\documentclass[aps,prd,nofootinbib]{revtex4-2}
\usepackage[colorlinks=true, pdfstartview=FitV, linkcolor=blue, citecolor=red, urlcolor=magenta]{hyperref}
\usepackage{graphicx}
\usepackage{latexsym}
\usepackage{amsmath}
\usepackage{amsfonts}
\usepackage{amssymb}
\usepackage{verbatim}
\usepackage{braket}

\newcommand{\be}{\begin{equation}}
\newcommand{\ee}{\end{equation}}
\newcommand{\bea}{\begin{eqnarray}}
\newcommand{\eea}{\end{eqnarray}}

\newcommand{\bb}{\bibitem}

\def\bb{\bibitem}

\def\bb{\bibitem}
\newcommand{\ben}{\begin{eqnarray}}
\newcommand{\een}{\end{eqnarray}}

\begin{document}

\title{Radiatively induced finite and (un)determined Chern-Simons-like terms
}


\author{$^{1}$E. Passos}
\email{passos@df.ufcg.edu.br}

\author{$^{1,2}$K. E. L. de Farias}
\email{klecio.lima@uaf.ufcg.edu.br}

\author{$^{1}$M. A. Anacleto}
\email{anacleto@df.ufcg.edu.br}

\author{$^{1,4}$E. Maciel}
\email{eugenio.maciel@df.ufcg.edu.br}

\author{$^{3}$C.A.G Almeida}
\email{cvniro@gmail.com}


\affiliation{$^{1}$Departamento de F\'{\i}sica, Universidade Federal de Campina Grande,\\
Caixa Postal 10071, 58429-900, Campina Grande, PB, Brazil.}
\affiliation{$^{2}$Departamento de F\' isica, Universidade Federal da Para\' iba,\\  Caixa Postal 5008, Jo\~ ao Pessoa, PB, Brazil.}
\affiliation{$^{3}$Departamento de Ci\^encias Exatas, Universidade Federal da Para\'iba,
 Rio Tinto, PB, Brazil}
\affiliation{$^{4}$Departamento de F\' isica-CCT,  Universidade Estadual da Para\' iba, Juv\^encio Arruda S/N, Campina Grande, PB, Brazil}



\begin{abstract}
The problem of Chern-Simons-like term induction via
quantum corrections in four-dimensions is investigated in two
different cases. In the first case, we consider two distinct
approaches to deal with the exact fermion propagator of the extended
QED theory up to the first order in the $b$-coefficient. We find
different results for distinct approaches in the same regularization
scheme. In the second case, we show that when we use a modified
derivative expansion method and another regularization scheme, we
obtain a result that exactly coincides with one of the results
obtained in the former case. This seems to imply an ambiguity
absence as one treats the fermion propagator and the self-energy
tensor properly.

\end{abstract}
 \maketitle

\vspace{-0.5cm}
\begin{center}
\end{center}

\section{Introduction}

The induction of the Chern-Simons-like Lorentz and CPT violating term, given by $\mathcal{L}_{C S}=$ $\frac{1}{2} k_\mu \epsilon^{\mu \nu \lambda \theta} F_{\nu \lambda} A_\theta$, being $k_\mu$ a constant vector characterizing the preferred direction of the space-time, is one of the most important results in the study of Lorentz symmetry violation $[1,2]$. This term which is known to have some important implications, such as birefringence of light in the vacuum [3], naturally emerges as a quantum correction in the theory suggested in [2] as a possible extension of QED:
$$
\mathcal{L}_{Q E D}=\bar{\psi}(i \not \partial-m) \psi-\bar{\psi} \psi \gamma_5 \psi-e \bar{\psi} A \psi
$$
where $b_\mu$ is a parameter introducing CPT symmetry breaking. Carrying out the integration over fermions, the relation between the coefficients $k_\mu$ and $b_\mu$ could be obtained in terms of some loop integrals with some of them being divergent. Therefore one has to implement some regularization scheme to calculate these integrals. Thus, the constant relating the coefficient $k_\mu$ and $b_\mu$ turns out to be dependent on the regularization scheme used $[4,5]$. Such dependence on the regularization scheme has been intensively discussed in $[6,7,8]$. However, there is an alternative study with absence of ambiguities put forward recently in $[9]$.

Based on the theory (1.1), the purpose of our study is to investigate different possibilities of finding ambiguities inherent to generation of Chern-Simons-like term via quantum corrections in four dimensions. We do this by using derivative expansion method of the fermion determinant $[10]$ and the imaginary time formalism.

The work structure is organized as follows: In section 2 we will investigate the induction of the Chern-Simons-like term via quantum corrections by using distinct approaches to deal with the exact fermion propagator. We find distinct relations between the coefficients $k_\mu$ and $b_\mu$ for different approaches, but with a same regularization scheme. Therefore, we conclude that different approaches to deal with the exact fermion propagator of the theory,

leads to a new ambiguity for the problem of radiatively induced Chern-Simons-like term. In section 3 we develop another method to investigate the present issue. By modifying the derivative expansion method, we obtain a different self-energy tensor. We then use a specific regularization scheme to find a finite result identical to that obtained in section 2 by using another regularization scheme. This effect seems to imply ambiguity absence in our calculations. Finally, in section 4 we present our conclusions.

\section{Inducing Chern-Simons-like term: Two different approaches}

In this section, we focus on the induction of the Chern-Simons-like term coefficient by expanding the self-energy (2.6) and using two distinct approaches to deal with the exact fermion propagator (2.7) up to the leading order in $b$. We find a new ambiguity because two different results appear.

The one-loop effective action $S_{e f f}[b, A]$ of the gauge field $A_\mu$ related to theory (1.1), can be expressed in the form of the following functional trace:
$$
S_{e f f}[b, A]=-i \operatorname{Tr} \ln \left(\not p-m-\not \beta \gamma_5-e \not A\right)
$$
This functional trace can be represented as $S_{e f f}[b, A]=S_{e f f}[b]+S_{e f f}^{\prime}[b, A]$, where the first term $S_{\text {eff }}[b]=-i \operatorname{Tr} \ln \left(\not p-m-\not \gamma_5\right)$ does not depend on the gauge field. The only nontrivial dynamics is concentrated in the second term $S_{\text {eff }}^{\prime}[b, A]$, which is given by the following power series:
$$
S_{e f f}^{\prime}[b, A]=i \operatorname{Tr} \sum_{n=1}^{\infty} \frac{1}{n}\left[\frac{1}{\not p-m-\not p \gamma_5} e \not A\right]^n .
$$
To obtain the Chern-Simons-like term we should expand this expression up to the second order in the gauge field:
$$
S_{e f f}^{\prime}[b, A]=S_{e f f}^{(2)}[b, A]+\ldots
$$
The dots in (2.3) stand for higher order terms in the gauge field. Here
$$
S_{\text {eff }}^{(2)}[b, A]=\frac{i e^2}{2} \operatorname{Tr}\left[\frac{1}{\not p-m-\not \beta \gamma_5} A \mathcal{1} \frac{1}{\not-m-\not p \gamma_5} A\right] .
$$
Using the derivative expansion method [10] one can find that the one-loop contribution to $S_{\text {eff }}^{(2)}[b, A]$ reads
$$
S_{\mathrm{eff}}^{(2)}[b, A(x)]=\frac{1}{2} \int d^4 x \Pi^{\alpha \mu \nu} F_{\alpha \mu} A_\nu
$$
where the one-loop self-energy $\Pi^{\alpha \mu \nu}$ is given by
$$
\Pi^{\alpha \mu \nu}=-\frac{i e^2}{2} \int \frac{d^4 p}{(2 \pi)^4} \operatorname{tr}\left[S_b(p) \gamma^\mu S_b(p) \gamma^\alpha S_b(p) \gamma^\nu\right],
$$
where
$$
S_b(p)=\frac{i}{\not p-m-\not p \gamma_5}
$$
is a $b^\mu$ dependent exact fermion propagator of the theory.

\subsection{ Approach I: Fermion propagator rationalized}

Firstly, we use the approximation developed in [11], where the exact propagator (2.7) is rationalized in the form
$$
S_b(p)=i\left[\frac{\not p+m-\gamma_5 \not p}{\left(p^2-m^2\right)}-\frac{2 \gamma_5(m \not p-(b \cdot p))(\not p+m)}{\left(p^2-m^2\right)^2}\right]+\cdots .
$$
Substituting (2.8) into (2.6), we can calculate the trace of gamma matrices, resulting in the following expression for the self-energy tensor [4]:
$$
\begin{aligned}
\Pi_{\mathbf{r}}^{\mu \alpha \nu}= & -2 i e^2 \int \frac{d^4 p}{(2 \pi)^4} \frac{1}{\left(p^2+m^2\right)^3}\left\{3 \varepsilon^{\alpha \mu \nu \theta}\left[b_\theta\left(p^2-m^2\right)-2 p_\theta(b \cdot p)\right]\right. \\
& \left.-2 b_\theta\left[\varepsilon^{\beta \mu \nu \theta} p_\beta p^\alpha+\varepsilon^{\alpha \beta \nu \theta} p_\beta p^\mu+\varepsilon^{\alpha \mu \beta \theta} p_\beta p^\nu\right]\right\},
\end{aligned}
$$
where $\Pi_{\mathbf{r}}^{\mu \alpha \nu}$ means self-energy tensor is rationalized up to the first order in $b$-coefficient. In Eq.(2.9), we turn the Minkowski space to an Euclidean space by performing the Wick rotation $x_0 \rightarrow-i x_0, p_0 \rightarrow i p_0, b_0 \rightarrow i b_0, d^4 x \rightarrow-i d^4 x$ and $d^4 p \rightarrow i d^4 p$. Note that by power counting, the momentum integral in (2.9) involves a finite term and terms with logarithmic divergences. In order to regularize such divergence, we use a scheme which we implement translation only on space coordinates of the momentum $p_\rho[12]$. Hence, we have
$$
p_\rho \rightarrow \vec{p}_\rho+p_0 \delta_{0 \rho}
$$
We use the covariance under spatial rotations which allows us to carry out the following replacement
$$
\vec{p}_\rho \vec{p}^\sigma \rightarrow \frac{\vec{p}^2}{D}\left(\delta_\rho^\sigma-\delta_{\rho 0} \delta_0^\sigma\right)
$$
Thus,
$$
\begin{aligned}
& 2 p_\rho(b \cdot p) \rightarrow 2\left(b_\rho \frac{\vec{p}^2}{D}-b_0 \delta_{\rho 0}\left(\frac{\vec{p}^2}{D}-p_0^2\right)\right), \\
& 2 p_\beta p^\alpha \rightarrow 2\left(\delta_\beta^\alpha \frac{\vec{p}^2}{D}-\delta_{\beta 0} \delta_0^\alpha\left(\frac{\vec{p}^2}{D}-p_0^2\right)\right) .
\end{aligned}
$$
We have that only the terms above can contribute to the Chern-Simons structure. Therefore, we can split the expression (2.4) into a sum of two parts, "covariant" and "noncovariant", i.e.,
$$
S_{\mathrm{eff}}^{\mathrm{cov}}=\frac{(-i)}{2} \int d^4 x I_1 \varepsilon^{\alpha \mu \nu \beta} b_\beta F_{\alpha \mu} A_\nu
$$
with
$$
I_1=\frac{-3 i e^2}{2 \pi} \int \frac{d^D \vec{p}}{(2 \pi)^D} \int_{-\infty}^{+\infty} d p_0\left[\frac{\left(1-\frac{4}{D}\right) \vec{p}^2+p_0^2-m^2}{\left(\vec{p}^2+p_0^2+m^2\right)^3}\right]
$$

and
$$
\begin{aligned}
S_{\mathrm{eff}}^{\mathrm{ncv}}= & \frac{i}{2} \int d^4 x I_2\left[3 \varepsilon^{\alpha \mu \nu 0} b_0 F_{\alpha \mu} A_\nu+b_\theta\left(\varepsilon^{0 \mu \nu \theta} F_{0 \mu} A_\nu\right.\right. \\
& \left.\left.+\varepsilon^{\alpha 0 \nu \theta} F_{\alpha 0} A_\nu+\varepsilon^{\alpha \mu 0 \theta} F_{\alpha \mu} A_0\right)\right],
\end{aligned}
$$
with
$$
I_2=\frac{i e^2}{\pi} \int \frac{d^D \vec{p}}{(2 \pi)^D} \int_{-\infty}^{+\infty} d p_0\left[\frac{\vec{p}^2-p_0^2}{\left(\vec{p}^2+p_0^2+m^2\right)^3}\right] .
$$
The integrals $I_1$ and $I_2$ over $p_0$ are finite and can be calculated by residues theorem. Hence, we have
$$
I_1=\frac{3 i e^2}{8 D} \int \frac{d^D \vec{p}}{(2 \pi)^D}\left[\frac{2(3-D) \vec{p}^2+D m^2}{\left(\vec{p}^2+m^2\right)^{5 / 2}}\right]
$$
and
$$
I_2=-\frac{i e^2}{8 D} \int \frac{d^D \vec{p}}{(2 \pi)^D}\left[\frac{(3-D) \vec{p}^2-D m^2}{\left(\vec{p}^2+m^2\right)^{5 / 2}}\right]
$$
Now, we can integrate over the spatial momentum in $D$-dimensions [15], and we find,
$$
I_1=\frac{3 i e^2}{8(4 \pi)^{D / 2}} \frac{\Gamma\left(1+\frac{\epsilon}{2}\right)}{\Gamma\left(\frac{5}{2}\right)\left(m^2\right)^{\epsilon / 2}}
$$
where $\epsilon=3-D$. The integral $I_2$ presents a value identically equal to zero. Therefore, the effective action Eq.(2.13) in the limit $D=3$ can be written in the form
$$
S_{\text {eff }}^{\text {cov }}=\frac{1}{2} \int d^4 x \varepsilon^{\alpha \mu \nu \beta} k_\beta F_{\alpha \mu} A_\nu
$$
where
$$
k_\beta=\frac{e^2}{16 \pi^2} b_\beta
$$
that coincides with the resulted of work [16] in the Schwinger constant field approximation.

\subsection{ Approach II: Fermion propagator expansion}

Now we use the approximation developed in [13] to expand the exact propagator (2.7) up to the first order in $b$-coefficient
$$
S_b(p)=\frac{i}{\not p-m}+\frac{i}{\not p-m}\left(-i p \gamma \gamma_5\right) \frac{i}{\not p-m}+\cdots
$$
In this case, we have our self-energy tensor in the form $[5,17]$ :
$$
\begin{aligned}
\Pi_{\mathbf{e}}^{\mu \alpha \nu}= & -2 i e^2 \epsilon^{\mu \alpha \nu \rho} \int \frac{d^4 p}{(2 \pi)^4} \frac{1}{\left(p^2+m^2\right)^3} \\
& \times\left[b_\rho\left(p^2-3 m^2\right)-4 p_\rho(b \cdot p)\right],
\end{aligned}
$$

where $\Pi_{\mathrm{e}}^{\mu \alpha \nu}$ means the expanded self-energy tensor. In Eq.(2.23), we also change the Minkowski space to Euclidean space. Note that by power counting, the integral in the momentum space Eq.(2.23) also present a finite term and another that diverges logarithmically. Thus, we use the same regularization scheme adopted in the rationalized fermion propagator approach. In this way, we obtain the following effective action in the limit of $D=3:$
$$
\tilde{S}_{\mathrm{eff}}^{\text {cov }}=\frac{1}{2} \int d^4 x \varepsilon^{\alpha \mu \nu \beta} \tilde{k}_\beta F_{\alpha \mu} A_\nu,
$$
where
$$
\tilde{k}_\beta=\frac{e^2}{4 \pi^2} b_\beta .
$$
Here, we also have the absence of "noncovariant" part. The result (2.25) is equivalent to the obtained in [18] where one uses a physical cutoff for fermions. According to the results (2.21) and (2.25), we found that the use of different approaches to deal with the exact fermion propagator leads to distinct relations between the coefficient $k_\mu$ and $b_\mu$ for the Chern-Simons-like term. Our results establish the following relation:
$$
\tilde{k}_\beta=k_\beta+\frac{3 e^2}{16 \pi^2} b_\beta
$$
that means
$$
\Delta k_\beta=\frac{3 e^2}{16 \pi^2} b_\beta
$$
The result (2.27) corresponds to variation between the results (2.21) and (2.25), that is identical to the result of Ref. [17] originated from the self-energy (2.23), in another regularization scheme. The same result has been found in the literature [13, 14].

\section{ Other aspects on induced Chern-Simons-like term}

In this section, we present an alternative method to compute induced Chern-Simons-like term, that is independent of the approach used to deal with the exact fermion propagator.
Let us rewrite the Eq.(2.2) in the form,
$$
S_{e f f}^{\prime}[b, A]=i \operatorname{Tr} \sum_{n=1}^{\infty} \frac{1}{n}\left[\frac{1}{\not p-m-e \not A} \not \gamma_5\right]^n .
$$
To obtain the Chern-Simons-like term we should expand this expression up to the leading order in $b$. Thus, for $n=1$, we have
$$
S_{e f f}^{(1)}[b, A]=i \operatorname{Tr}\left[\frac{1}{\not p-m-e A} \not \gamma_5\right] .
$$
Using the relation
$$
\frac{1}{A-B}=\frac{1}{A}+\frac{1}{A} B \frac{1}{A}+\frac{1}{A} B \frac{1}{A} B \frac{1}{A}+\cdots
$$

for $A=\not p-m$ and $B=e A$, we find
$$
S_{e f f}^{(1)}[b, A]=i e^2 \operatorname{Tr}\left[\frac{1}{p-m} \not \gamma_5 \frac{1}{p-m} A \frac{1}{p-m} A\right],
$$
where one was considered the cyclic property of the trace in the product of the $\gamma$-matrices. By using derivative expansion method, we find the following effective action:
$$
S_{\text {eff }}^{(1)}[b, A(x)]=\frac{1}{2} \int d^4 x \Pi^{\alpha \mu \nu} F_{\alpha \mu} A_\nu,
$$
where the one-loop self-energy $\Pi^{\alpha \mu \nu}$ is given by
$$
\begin{aligned}
\Pi^{\mu \alpha \nu}= & -2 i e^2 \int \frac{d^4 p}{(2 \pi)^4} \frac{1}{\left(p^2-m^2\right)^3}\left\{\varepsilon^{\alpha \mu \nu \theta} b_\theta\left(p^2-m^2\right)\right. \\
& \left.-2 b_\theta\left[\varepsilon^{\alpha \nu \theta \beta} p_\beta p^\mu-\varepsilon^{\alpha \mu \theta \beta} p_\beta p^\nu\right]\right\} .
\end{aligned}
$$
Note that in the self-energy tensor (3.6) there exists a convergent contribution and the remaining term diverges logarithmically. However, differently of the previous situations, in this case the calculation of the divergent integrals is very delicate. Calculating this self-energy tensor by using the same regularization scheme of the section 2 , we find a null result, and in turn, the absence of the Chern-Simons term. On the other hand, there also exists the possibility of using in Eq.(3.6) the relation
$$
\int \frac{d^4 p}{(2 \pi)^4} p_\mu p_\nu f\left(p^2\right)=\frac{g_{\mu \nu}}{4} \int \frac{d^4 p}{(2 \pi)^4} p^2 f\left(p^2\right),
$$
that naturally removes the logarithmic divergence. As a result, we have only the finite contribution
$$
\Pi^{\mu \alpha \nu}=\frac{e^2}{16 \pi^2} \varepsilon^{\mu \alpha \nu \theta} b_\theta
$$
In this case, we find
$$
S_{\mathrm{eff}}^{(1)}[b, A(x)]=\frac{1}{2} \int d^4 x k_\beta \varepsilon^{\mu \alpha \nu \beta} \partial_\alpha A_\mu A_\nu
$$
where
$$
k_\beta=\frac{e^2}{16 \pi^2} b_\beta
$$
Note that the result (3.10) is the same as the result found in (2.21) where the exact fermion propagator was rationalized up to the first order in the $b$-coefficient. Thus, we have here another surprising effect: The result $k_\beta=\frac{e^2}{16 \pi^2} b_\beta$ given in (2.21) as result of dimensional regularization and also given in (3.10) as result of the Lorentz preserving regularization (3.7), appears to be independent of the regularization scheme used by properly carrying out the self-energy tensor. The result (3.10) was also obtained in the work [14] by using massless exact fermion propagator. We observe that the factor $\left(e^2 / 16 \pi^2\right)$, is exactly the same as found in the well-known Adler-Bell-Jackiw anomaly $[19,20,21]$.
\section{Conclusion}\label{sec5}

We have investigated the induction of Chern-Simons-like term via quantum corrections in two different situations. Firstly, we use a same regularization to different approaches to deal with the exact fermion propagators up to the leading order in the $b$-coefficient. In this case, our results are finite and agree with other results in the literature, but they do not agree with each other, because they are different depending on the approach used. Moreover, it generates values whose difference is exactly: $\Delta k_\beta=\left(3 e^2 / 16 \pi^2\right) b_\beta$, which agrees with the result of Ref. [17] found in another context. We conclude that this is due to the different approximation of the exact fermion propagator of theory. The problem was also investigated in a context independent of the approaches used to deal with the exact fermion propagator. In this case, we modify the derivative expansion method and obtain a new self-energy tensor for the effective action. The momentum integrals were calculated by using another regularization scheme. As a result, we obtained a relation between the coefficients $k_\beta$ and $b_\beta$ identical to the one obtained in the case where it was used the fermion propagator expansion. We also observed that the parameter of proportionality between these two coefficients is exactly the same as the one found in the well-known Adler-Bell-Jackiw anomaly. In our calculations, we also observed that the "noncovariant" contributions for the Chern-Simons-like term are absent, as was anticipated in [4] in the finite temperature context. Therefore, we insist that a complete comprehension of these question will require further investigations.

{\acknowledgments} We would like to thank CNPq, CAPES and CNPq/PRONEX/ FAPESQ-PB (Grant nos. 165/2018 and 015/2019), for partial financial support. MAA and EP acknowledge support from CNPq (Grant nos. 306398/2021-4, 312104/2018-9, 304290/2020-3). KELF thanks financial support from the Brazilian agency CNPq.



\end{document}